\documentclass[journal]{IEEEtran}

\ifCLASSINFOpdf

\else

\fi

\usepackage{amssymb}
\usepackage{xcolor}
\usepackage{cite}
\usepackage{multirow}
\usepackage{booktabs}
\usepackage{makecell}
\usepackage{amsmath}
\usepackage{textcomp}
\usepackage{bm}
\usepackage{color}
\usepackage{setspace}
\usepackage{algorithm}
\usepackage{amsmath}
\usepackage{algorithmic}
\usepackage{amsthm,amsmath,amssymb}
\usepackage{mathrsfs}
\usepackage{graphics}
\usepackage{array,diagbox,siunitx}
\usepackage{graphicx}
\usepackage{subfigure}
\usepackage{url}
\usepackage{hyperref}
\usepackage[hyphenbreaks]{breakurl}

\usepackage{epsfig}
\usepackage{amssymb}

\usepackage{algorithm}
\usepackage{algorithmic}

\usepackage{lineno,hyperref}
\modulolinenumbers[5]

\begin{document}

\title{Transceiver Cooperative Learning-aided Semantic Communications Against Mismatched Background Knowledge Bases}

\author{Yanhu Wang, \emph{Student Member}, \emph{IEEE}, and~Shuaishuai Guo, \emph{Senior Member}, \emph{IEEE}
\thanks{Y. Wang and S. Guo are all  with the School of Control Science and Engineering, Shandong University, China, and also with Shandong Key Laboratory of Wireless Communication Technologies, Shandong University, China (e-mail: yh-wang@mail.sdu.edu.cn; shuaishuai$\_$guo@sdu.edu.cn). }}


\maketitle

\begin{abstract}

Semantic communications learned on background knowledge bases (KBs) have been identified as a promising technology for communications between intelligent agents. Existing works assume that transceivers of semantic communications share the same KB. However, intelligent transceivers may suffer from the communication burden or worry about privacy leakage to exchange data in KBs. Besides, the transceivers may independently learn from the environment and dynamically update their KBs, leading to timely sharing of the KBs infeasible. All these cause the mismatch between the KBs, which may result in a semantic-level misunderstanding on the receiver side.
To address this issue, we propose a transceiver cooperative learning-assisted semantic communication (TCL-SC) scheme against mismatched KBs. In TCL-SC, the transceivers cooperatively train semantic encoder and decoder neuron networks (NNs) of the same structure based on their own KBs. They periodically share the parameters of NNs. To reduce the communication overhead of parameter sharing, parameter quantization is adopted. Moreover, we discuss the impacts of the number of communication rounds on the performance of semantic communication systems.
Experiments on real-world data demonstrate that our proposed TCL-SC can reduce the semantic-level misunderstanding on the receiver side caused by the mismatch between the KBs, especially at the low
signal-to-noise (SNR) ratio regime.
\end{abstract}




\IEEEpeerreviewmaketitle

\section{Introduction}

Due to tremendous breakthroughs in artificial intelligence (AI) and powerful chipsets, many intelligent applications such as Sari of the Apple company, and self-driving vehicles, have sprung up. Semantic communications, as a communication paradigm beyond transmitting bits\cite{Qin2021Survey,GuoVTC}, aim to precisely convey the meaning of messages, rather than accurately transmitting each symbol.
It can significantly reduce data traffic and meanwhile well support the communication requirement of intelligent agents. It has been recognized as a promising technology to make wireless networks significantly more intelligent, energy-efficient, and sustainable.

Owing to advances in deep learning, in particular natural language processing (NLP) and computer vision, digging the semantic meaning of data for transmission becomes possible. In recent years, semantic communication systems learned on background knowledge bases (KBs) at transceivers have been developed for delivering text \cite{farsad2018deep,Xie2020, jiang2022deep}, image \cite{bourtsoulatze2019deep,ShaoZ20}, speech \cite{Weng2021}, as well as multimodal data \cite{zhang2022unified}.
In semantic communication systems, the transmitter uses a semantic encoding module to extract semantic information based on its own KB, and the receiver uses a semantic decoding module to recover the meaning of messages based on its own KB.
To make the transceivers have the same interpretation of the transmitted semantic data,  existing works, e.g., \cite{Xie2020,farsad2018deep,jiang2022deep, bourtsoulatze2019deep,ShaoZ20, Weng2021,zhang2022unified}, assume the transceivers have the same KBs.
However, in practice, the KBs of the transmitter and receiver may be the same initially, they may become different due to the variations of the environment and/or the strength of the device's ability to acquire data.
The mismatch between the KBs of the transmitter and receiver can cause misunderstanding on the receiver side.
For instance, the KB of the transmitter has knowledge about cars, ships, and birds, while the KB of the receiver has knowledge about cars, ships, and planes.
When the transmitter transmits semantic information about birds, the receiver cannot understand it precisely.
This calls for a way to address the problem of the mismatch between the KBs and avoid misunderstanding at the receiver.

The communication participant sharing their own data with another participant to update the KB seems to be a viable approach.
However, there are the following intractable challenges when applying it to address the above problem:
\begin{itemize}
    \item
    \textbf{Massive communication overhead}:
    In reality, the data acquired by devices are mainly images and videos.
    Direct sharing of these data to another communication participant may result in unaffordable communication overhead.
    \item
   \textbf{Expensive training cost}:
    In general, the storage capacity and computing power of the device are limited.
    Shared data, especially images and videos, undoubtedly increase the storage cost and training cost of the device.
    \item
    \textbf{Privacy concerns}:
    Data shared during the communication may be illegally stolen by adversaries leading to privacy leakage.
    Also out of privacy protection, some participants are reluctant to share the acquired data with another communication participant.
\end{itemize}
These challenges motivate us to propose a feasible approach to address the mismatch between the KBs in semantic communication systems.

\begin{figure*}
       \centering
       \includegraphics[width=1\linewidth]{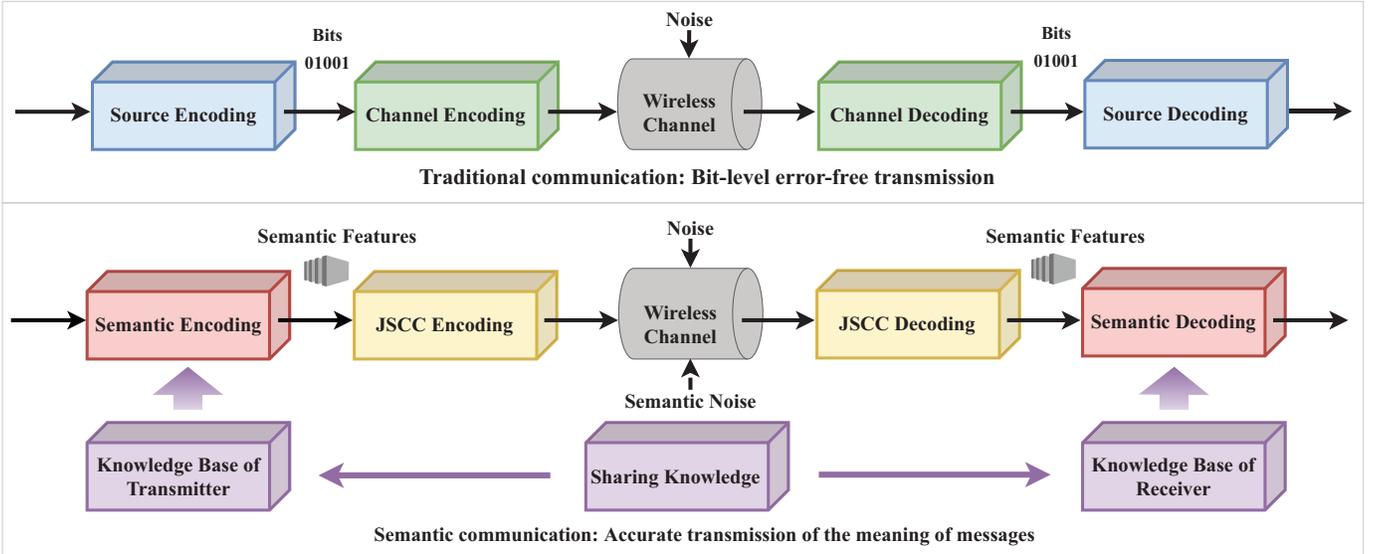}
       \caption{A comparison between traditional communication system and deep learning-enabled semantic communication system. The upper region of the graph corresponds to the traditional communication system, and the lower region corresponds to the semantic communication system. In the traditional communication system, the source is encoded as bit sequences by source encoding and recovered by source decoding module at the receiver. In the semantic communication system, semantic encoding module is fused to extract semantic features. In addition to being disturbed by physical noise, semantic communication is also disturbed by semantic noise which is caused by the ambiguity of words, sentences or symbols between the transmitter and receiver.}
       \label{TCMSCM}
\end{figure*}

In this article, a new semantic communication framework is developed to combat the mismatched KBs of the transmitter and receiver without causing excessive communication overhead, training costs as well as privacy concerns.
As indicated in previous studies\cite{shlezinger2020uveqfed}, only exchanging parameters of neuron networks (NNs) without data exchange could achieve a balance between data privacy protection and data sharing computing.
Inspired by these findings,
we design a transceiver cooperative learning-aided semantic communication (TCL-SC) framework, in which the transceivers cooperatively train the semantic encoder
and decoder NNs of the same structure based on their own KBs and periodically share the parameters of NNs.
Obviously, TCL-SC still needs a large amount of communication overhead for exchanging the parameters precisely.
Inspired by previous work \cite{xie2020lite},  quantization is adopted to reduce communication overhead.
Also, the impact of the number of communication rounds on the performance of semantic communications is investigated, trying to find out the appropriate value that can reduce the communication overhead without reducing the performance.


The main contributions of this article can be summarized as follows.
\begin{itemize}
    \item
    A TCL-SC scheme is proposed by using parameter sharing to minimize the semantic loss caused by mismatched KBs, which avoids excessive costs as data sharing needs.
    \item
    To reduce the communication overhead caused by parameter sharing, parameter quantization is adopted to reduce the weights resolution. The effect of the number of communication rounds on TCL-SC is investigated, and a suitable value is chosen to achieve a compromise between communication overhead and performance.
    \item
    The effectiveness of the proposed TCL-SC is validated on text transmission tasks.
    Extensive experiments demonstrate that, compared with baselines, the proposed TCL-SC can well combat mismatched KBs and reduce the misunderstanding on the receiver side, especially at the low signal-to-noise (SNR) ratio.
\end{itemize}

In the remainder of the article, deep learning-enabled semantic communications are first introduced, and the difference from traditional communication systems is emphasized.
Next, the TCL-aided semantic communication scheme against mismatched KBs is illustrated.
Then, comprehensive experiments are conducted are concluded and future directions are presented.
Finally, conclusions are drawn.

\section{Deep Learning-Enabled Semantic Communication System}

Recent advances in deep learning have made substantial progress in semantic communications, which has been dormant for many years\cite{Qin2021Survey,GuoVTC}.
As shown in Fig.~\ref{TCMSCM}, a DL-enabled semantic communication system consists of a transmitter and a receiver.
The transmitter includes a semantic coding module and a joint source and channel coding (JSCC) module, where the semantic encoding module based on the KB of the transmitter extracts semantic features from the raw data input, and the JSCC encoding module maps extracted semantic features to the channel input.
As the opposite process, the receiver is equipped with a JSCC decoding module and a semantic decoding module.
The JSCC decoding module decodes the channel output for mitigating the channel distortion and attenuation, then the semantic decoding module based on the KB of the receiver recovers it to the original messages.
According to the previous papers \cite{farsad2018deep, Xie2020, jiang2022deep,bourtsoulatze2019deep, ShaoZ20, Weng2021},
for text sources, the semantic encoding and semantic decoding modules can be composed of long short-term memory (LSTM)\cite{farsad2018deep} or Transformer\cite{Xie2020} networks. For image sources, convolutional neural networks\cite{bourtsoulatze2019deep,ShaoZ20} can be used to extract and recover semantic information. For speech sources, SE-ResNet networks\cite{Weng2021} can be used to construct the semantic encoding and semantic decoding module.
The JSCC encoding and decoding modules can generally be constructed by fully connected layers.

Compared to traditional communication systems which pursue bit-level error-free transmission, semantic communications aim an accurate transmission of the meaning of messages. The semantic encoding and decoding neuron networks (NNs) play the role of extracting and recovering the meaning of messages. For accuracy, they have to be trained on a KB.  The KB plays an important role in provisioning the accuracy of message meaning.
Existing works assume that transceivers of semantic communications share the same KB. However, intelligent transceivers may suffer from the communication burden or worry about privacy leakage to exchange data in KBs. Besides, the transceivers may independently learn from the environment and dynamically update their KBs, leading to timely sharing of the KBs infeasible. All these cause the mismatch between the KBs, which may result in a semantic-level misunderstanding on the receiver side.
Thus, it is important to develop a semantic communication scheme to mitigate the impact of the mismatch.

\begin{figure*}
       \centering
       \includegraphics[width=1\linewidth]{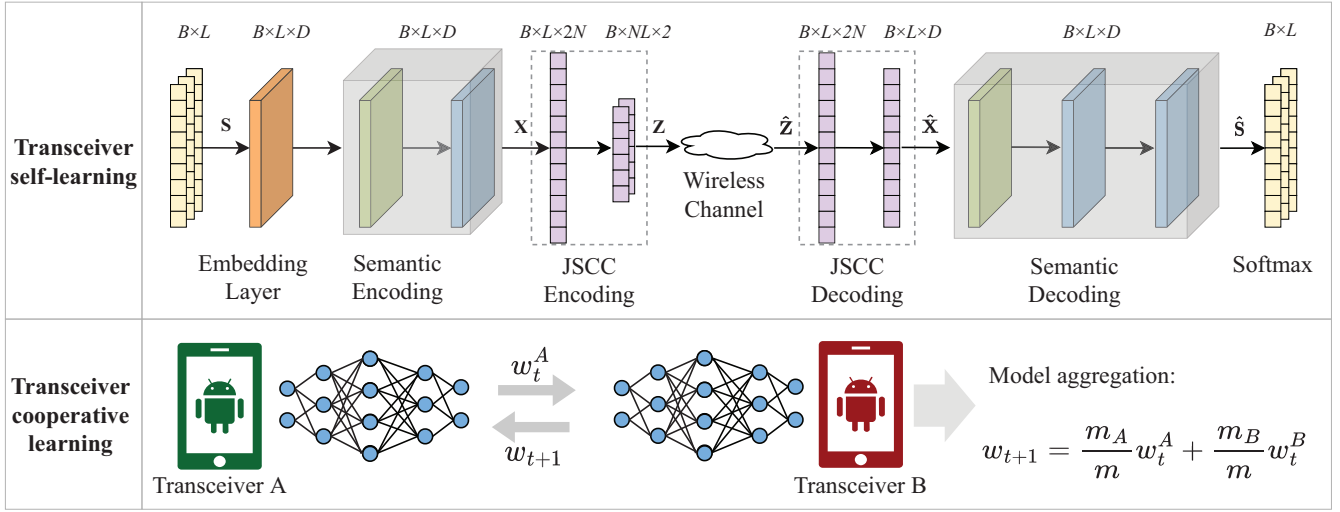}
       \caption{The framework of our proposed TCL-SC: (1) transceiver self-learning: the transceivers (e.g., A and B) train the semantic encoder and decoder of the same structure based on their own KBs. In order to reduce the communication overhead in cooperative learning, network parameter quantization is introduced to reduce the resolution of weights; (2) transceiver cooperative learning: the model parameters of transceivers A and B are aggregated on transceiver B. During model aggregation, the weight (e.g., $m_A$) assigned to the model is set according to the data size of the transceiver.}
       \label{TCLSC}
\end{figure*}

\section{Transceiver Cooperative Learning-aided Semantic Communication System for Text Transmission}

In this section,  a transceiver cooperative learning-aided semantic communication system to address the mismatch between KBs is illustrated.

\subsection{Transceiver Cooperative Learning-aided Semantic Communication}

The proposed TCL-SC is shown in Fig.~\ref{TCLSC}.
The transceivers cooperatively train the encoding and decoding NNs of the same structure based on their own KBs (i.e. transceiver self-learning) and periodically share the parameters of NNs (i.e. transceiver cooperative learning).

In the transceiver self-learning phase, the input data $\mathbf{S}\in\Re ^{B\times L}$ as a dense vector goes through the semantic encoding module, which extracts the semantic information $\mathbf{X}\in\Re ^{B\times L \times D}$, with $B$ being the batch size, $L$ representing the input data size, and $D$ stands for the output dimension of semantic encoding module.
The JSCC encoding module then maps $\mathbf{X}$ to the channel input symbols $\mathbf{Z}\in\Re ^{B\times NL \times 2}$.
The encoded features in $\mathbf{Z}$ are transmitted to the receiver over the wireless channel with noise, and the JSCC decoding module decodes the noise-corrupted features $\mathbf{\hat{Z}}\in\Re ^{B\times NL \times 2}$ to $\mathbf{\hat{X}}\in\Re ^{B\times L \times D}$.
Finally, the semantic decoding module recovers $\mathbf{\hat{X}}$ to $\mathbf{\hat{S}}\in\Re ^{B\times L \times D}$.
The whole network learns in an end-to-end manner. In the transceiver cooperative learning phase, one transceiver (e.g. A) shares the network parameters of the system model with the same architecture to the other transceiver (e.g. B), and the network parameters of both transceivers are aggregated on transceiver B. For simplicity, the weight assigned to the model aggregation is set to be propositional to the data sizes of the transceivers.
Transceiver B then sends the aggregated network parameters to transceiver A, and the network parameters of both transceivers are updated. 
The two phases alternate until the termination condition is met.

Note that, TCL-SC\footnote{Federated learning typically involves a large number of mobile devices that are connected to a central server. However, in the proposed TCL-SC, only two transceivers are considered.} could address the mismatch between KBs of the transmitter and receiver skillfully, but it also suffers from extra communication overhead. The communication overhead depends on the number of transmitted bits as well as the communication rounds. Therefore
to reduce the communication overhead, one can either the number of transmitted bits or that of the communication rounds. 

\subsection{Network Parameter Quantization}
As aforementioned, a feasible way is to convert the weights  of the trained NNs from high-precision to low-precision. The parameter compression can not only reduce the communication overhead of the  parameter sharing but also can improve the inference speed\cite{xie2020lite}.
In \cite{jacob2018quantization}, the authors proposed a quantization approach that quantizes both weights and activations from FP32 representation to INT8 representation.
The approach can reduce the model size without significantly degrading the model performance.
Thus, we use the approach proposed in \cite{jacob2018quantization} for the parameter sharing in TCL-SC.
This not only reduces the communication overhead, but also makes the proposed TCL-SC suitable for devices with limited resources.

\subsection{Impact of The Number of  Communication Rounds}

Intuitively, reducing the number of communication rounds can also reduce the communication overhead caused by sharing parameters of NNs.
However, the performance of the proposed TCL-SC scheme may severely degrade with the number of communication rounds. This is because the performance of semantic communication systems may fall into local optimality as the number of communication rounds decreases.
Hence, we expect  to find an appropriate value for achieving a compromise between communication overhead and performance.

\begin{figure}
       \centering
       \includegraphics[width=1\linewidth]{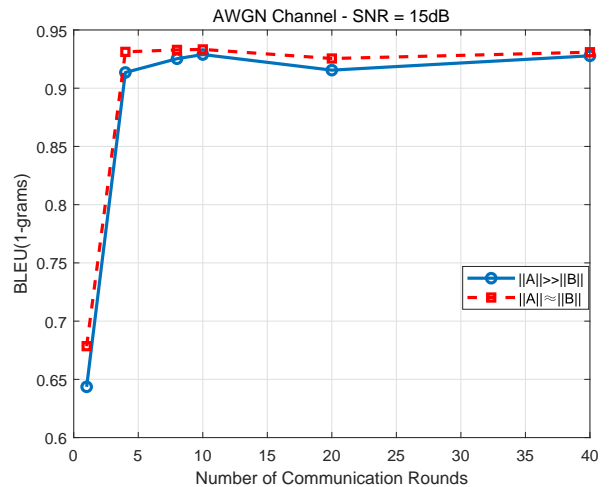}
       \caption{Impact of the number of communication rounds on the performance. Initially, the addition of communication rounds increases performance, but after a point, the gains stagnate.}
       \label{Round}
\end{figure}

 The impact of the number of communication rounds on the proposed TCL-SC performance is investigated in two cases.
\begin{itemize}
    \item
    \textbf{Case 1}: The data perceived by transceiver A is 
   different that perceived by transceiver B, but the data size is approximately equal, denoted as $\|A\|\approx\|B\|$.
    \item
    \textbf{Case 2}: The data perceived by transceiver A is 
   different that perceived by transceiver B, and the data size of transceiver A is much larger than that of transceiver B, denoted as $\|A\|\gg\|B\|$.
\end{itemize}
The impact of communication rounds is investigated by experiments. In this article, text transmission experiments are conducted. We select the
dataset \emph{proceedings of the European Parliament} \cite{Europarl2005}. The dataset contains around $2$ million sentences and $50$ million words.
We pre-process the dataset and select sentences consisting of $4-30$ words from it. In case 1, we divide the dataset into four parts: $10$\% as initial public data, $40$\% as private data of transceiver A, $40$\% as private data of transceiver B, and $10$\% as test data. In case 2, $60$\% is the private data of transceiver A, $20$\% is the private data of transceiver B, and the rest is unchanged.
The proposed TCL-SC scheme is trained over the additive white Gaussian noise (AWGN) channels with a signal-to-noise ratio (SNR) of $15$ dB with a cross-entropy loss function.
The total learning epochs of the transceivers is $80$, and the number of communication rounds is set to $1$, $4$, $8$, $10$, $20$, and $40$, respectively.
We adopt bilingual evaluation understudy (BLEU) score\cite{papineni2002bleu} as the performance metric.
The BLEU score measures the similarity between words, and its value ranges from $0$ to $1$.
The larger the value, the better the system performance.
The BLEU scores of the experiments  with different numbers of communication rounds are illustrated in Fig.~\ref{Round}. 

It is observed that the addition of communication rounds increases the performance initially in case 1.
However, with the increase of communication rounds to a certain number, the performance is no longer significantly improved.
In case 2, the BLEU score has shown roughly the same tendency.
Experiment results suggest that the number of communication rounds equal to 8 could provide a reasonable performance compromise.


\section{Performance Comparison and Discussions}

In this section, the effectiveness of the proposed TCL-SC is verified by comparing it with other semantic communication scheme and the traditional source coding and channel coding approach over the AWGN channels. The text transmission is chosen for the experiment. It is noteworthy the proposed TCL-SC scheme is not limited to the source type and can be applied to semantic communication systems with any source type.

\subsection{Experimental Setup}

\textbf{Dataset:} \emph{The proceedings of the European Parliament}\cite{Europarl2005} is adopted to demonstrate the performance comparison.
We pre-process the dataset and select sentences consisting of $4-30$ words from it.
Further, we divide the dataset into four parts: $10$\% as initial public data, $60$\% as private data of transceiver A, $20$\% as private data of transceiver B, and $10$\% as  test data.

\textbf{Network:}
For fair comparisons, the network structures of the transmitter and receiver in TCL-SC are the same as that of DeepSC \cite{Xie2020}, where the semantic encoding and decoding modules for extracting and recovering semantic information are composed of Transformer encoder and decoder.

\textbf{Baselines:} A DL-enabled semantic communication scheme and a traditional communication scheme are chosen as baselines.
\begin{itemize}
    \item
    DeepSC\cite{Xie2020}:
    DeepSC is a deep learning-based semantic communication system, which extracts semantic information from texts via the Transformer encoder and then maps semantic information to the channel input. The loss function of DeepSC is the cross-entropy.

    \item
    Separate source-channel coding scheme:
    We consider Huffman coding for source coding.
    As for channel coding, we use Reed-Solomon (RS) coding\cite{RS1960}. The 16 quadrature amplitude modulation (16-QAM) is adopted. It is noteworthy that the KB mismatch problem will not affect the separate source-channel coding scheme, because it aims to achieve perfect bit transmission.

\end{itemize}

\textbf{Metrics:} In addition to the BLEU score, we also adopt sentence similarity\cite{Xie2020} as performance metric.
The sentence similarity compares the difference between sentences.
Similar to the BLEU score, the sentence similarity ranges from $[0,1]$, and the larger the value, the better the performance of the communication system.


\subsection{Experimental Results}
\begin{figure}[htbp]
\centering
\vspace{-0.35cm}
\subfigtopskip=2pt
\subfigbottomskip=2pt
\subfigcapskip=-2pt
\subfigure[]{
\centering \includegraphics[width=1\linewidth, keepaspectratio=false]{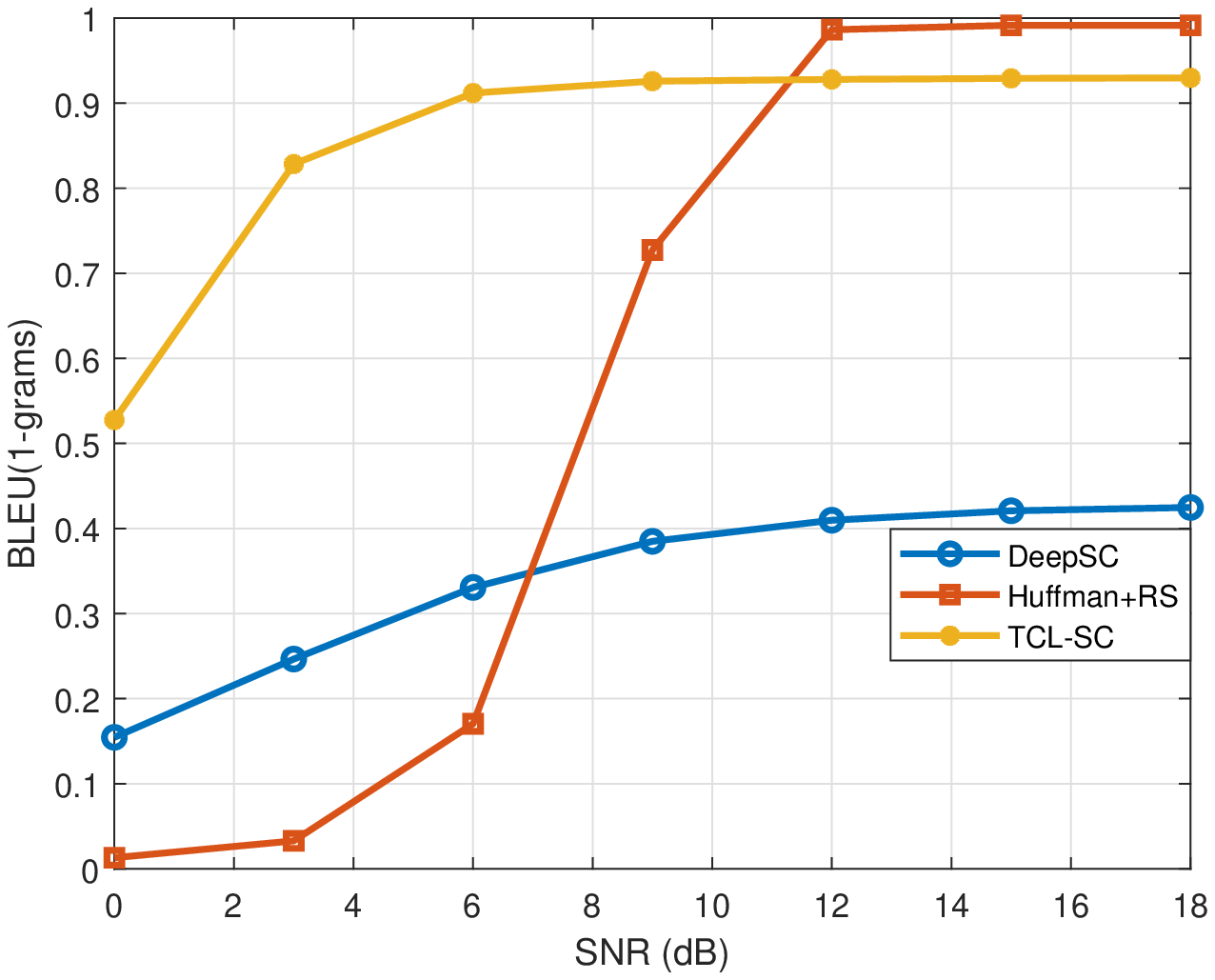}
}
\vspace{0.25cm}
\subfigure[]{
\centering \includegraphics[width=1\linewidth, keepaspectratio=false]{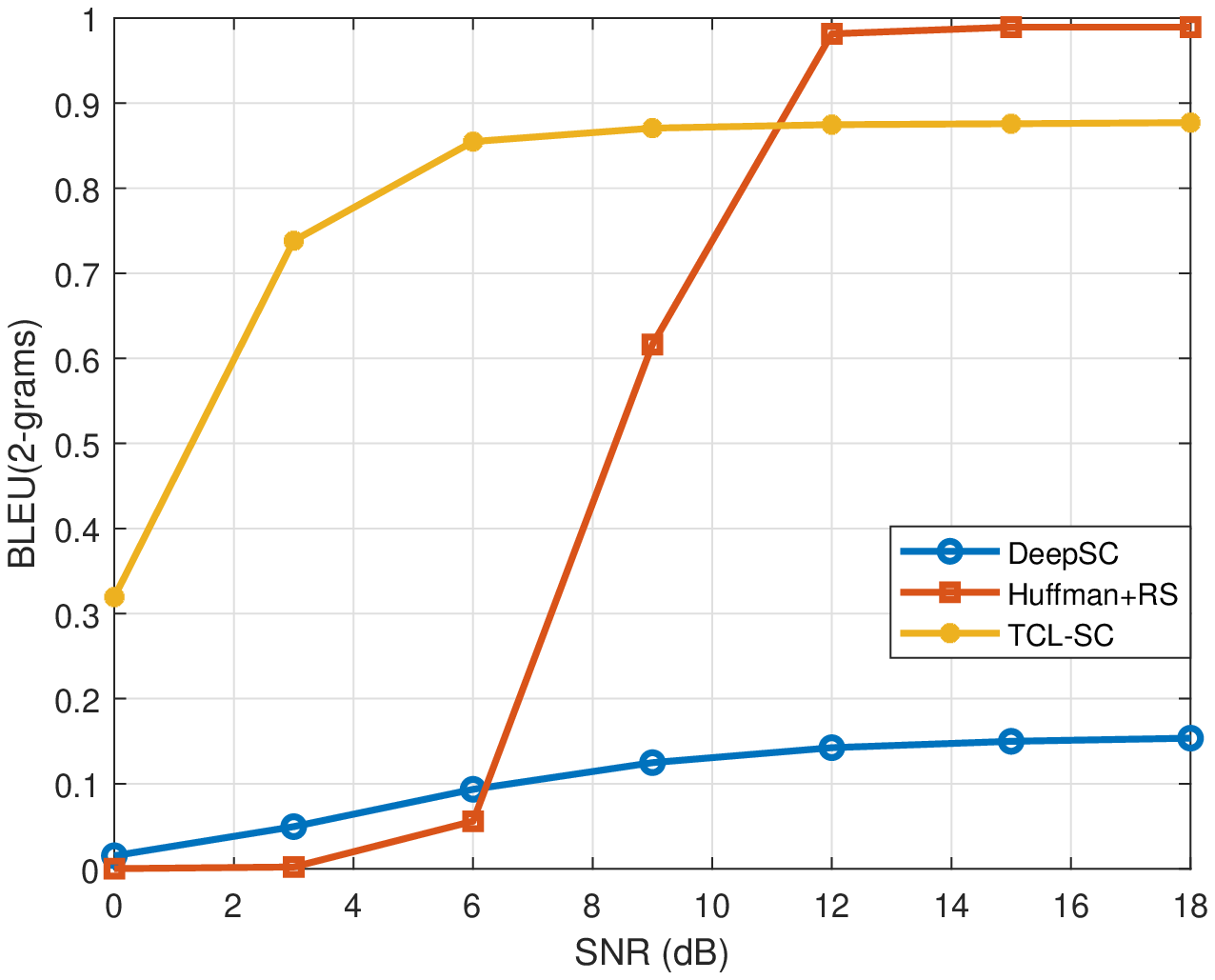}
}
\caption{BLEU score versus SNR, where both the DeepSC and TCL-SC are trained over the AWGN channels at an SNR of $15$ dB.}
\label{BLEU}
\end{figure}
With the same number of transmitted symbols, we test the BLEU score under different SNR over the AWGN channels, as shown in Fig.~\ref{BLEU}.
BLEU(1-gram) and BLEU(2-grams) calculated the 1-gram difference and 2-grams difference between the transmitted and recovered sentences, respectively.
Take the sentence ``the cat is on the sofa'' for example, 1-gram: ``the'',``is'',``on'',``the'', and ``sofa'', 2-gram: ``the cat'',``cat is'',``is on'',``on the'', and ``the sofa''.
Furthermore, take 2-grams as an example, if the recovered sentence contains the above phrases and the number is the same, then the BLEU score is $1$.

As can be seen from Fig.~\ref{BLEU}.(a), the BLEU(1-gram) score of the traditional communication scheme is higher than that of DeepSC and the proposed TCL-SC when the SNR is greater than $12$ dB. This is because channel impairments are relatively small. However, in the low SNR regime, the BLEU(1-gram) score of the proposed TCL-SC achieved is higher than that of the two baselines.
In particular, the performance of the proposed TCL-SC is much better than that of the DeepSC in both low and high SNR regimes.
In Fig.~\ref{BLEU}.(b), BLEU(2-grams) score obtained by each scheme decreased slightly but has shown roughly the same tendency with BLEU(1-gram) score.
These comparisons demonstrate that the existing semantic communication schemes cannot cope with the mismatched KBs, and the proposed TCL-SC could reduce the misunderstanding on the receiver side caused by the mismatched KBs.


\begin{figure}
       \centering
       \includegraphics[width=1\linewidth]{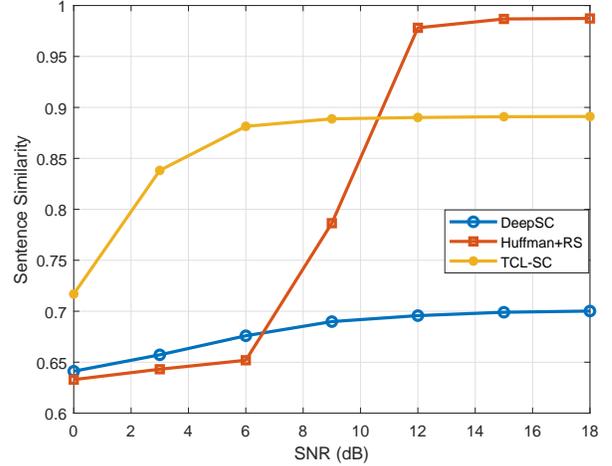}
       \caption{Sentence similarity versus SNR, where both the DeepSC and TCL-SC are trained over the AWGN channels at an SNR of $15$ dB.}
       \label{SS}
\end{figure}

Notice that, the BLEU score is still the pursuit of the accuracy of the symbol in essence, and it does not measure the performance of the communication system from the semantic level.
For example, ``sofa'' is recovered to ``lounges'', and the BLEU score will drop,
but in fact, the semantics have not changed.
In order to more appropriately measure the performance of semantic communication system, in Fig.~\ref{SS}, we show the relationship between sentence similarity and the SNR for each scheme over the AWGN channels.

It can be observed that, at the high SNR, the sentence similarity obtained by the traditional method is higher than that obtained by the proposed TCL-SC.
The sentence similarity obtained by DeepSC is much smaller than that obtained by the proposed TCL-SC.
At the low SNR, the sentence similarity obtained by the proposed TCL-SC is higher than that of the comparison schemes.
These comparisons further indicate that our proposed TCL-SC could well cope with the mismatched KBs in semantic communication systems, especially at the low SNR.

\section{Future Directions}

While the proposed TCL-SC has been demonstrated to be effective in combating mismatched KBs, there are still some issues worth further investigation.

\textbf{Imperfect transmission of NN parameters:}
As this article focuses on addressing the mismatch between the KBs, in the phase of transceiver cooperative learning, the transmission of the NN parameters is assumed to be perfect.
In practice, the channel for NN parameter transmission is typically imperfect. The noisy model parameter may greatly affect the performance of TCL-SC. Therefore, the impact of noisy parameter transmission on TCL-SC needs further investigation.

\textbf{Data importance-based weights setting:}
For simplicity, the weight in TCL-SC assigned to the model aggregation is set to be propositional to the data sizes of the transceivers.
The data importance that describes the contributions to model updates has been overlooked. In the future, the importance of data can be reflected in the weighted model aggregation.

\textbf{TCL-SC for multi-user communications:}
In this article, we only considered point-to-point semantic communication.
In practice, a single transceiver can usually communicate with multiple transceivers (e.g., transceiver A$\leftrightarrow$transceiver B, transceiver C$\leftrightarrow$transceiver A).
In the multi-transceiver communication scenario, it is much more complicated to adopt TCL-SC against the mismatched KBs.
Thus, more research efforts, such as how to avoid the semantic-level interference of multi-transceivers cooperative learning and how to optimize computing and communication resources, can be done on TCL-SC in multi-user communication scenarios in future work.

\section{Conclusions}

This article  proposed a TCL-SC scheme,
where the transceivers periodically share the network parameters of the semantic encoder and decoder instead of directly exchanging data.
In TCL-SC, the network parameter quantization is adopted to reduce the weights resolution, thus reducing the communication overhead caused by sharing the parameters of NNs.
Furthermore, we studied the effect of the number of communication rounds on TCL-SC.
Experiments of text transmission have demonstrated our proposed TCL-SC could well reduce the misunderstanding on the receiver side caused by the mismatched KBs, especially at the low SNR ratio regime. Future research directions regarding TCL-SC were discussed.

\ifCLASSOPTIONcaptionsoff
  \newpage
\fi

\bibliography{reference}
\vspace{-25 pt}

\end{document}